# *In situ* hydride breathing during the template-assisted electrodeposition of Pd nanowires

Giuseppe Abbondanza[1,2], Andrea Grespi[1,2], Alfred Larsson[1,2], Dmitry Dzhigaev[1,2], Lorena Glatthaar[3], Tim Weber[3], Malte Blankenburg[4], Zoltan Hegedüs[4], Ulrich Lienert[4], Herbert Over[3], Gary S. Harlow[5], Edvin Lundgren[1]

1 Division of Synchrotron Radiation Research, Lund University, Professorsgatan 1, Lund, Sweden

2 NanoLund, Lund University, Professorsgatan 1, Lund, Sweden

3 Institute of Physical Chemistry, Justus Liebig University, Heinrich-Buff-Ring 17, 35392 Giessen, Germany

4 Deutsches Elektronensynchrotron DESY, Notkestr. 85, 22607 Hamburg, Germany

5 Department of Chemistry and Biochemistry and the Oregon Center for Electrochemistry, University of Oregon, Eugene, OR 97403, USA

**Abstract**

We investigated the structural evolution of electrochemically fabricated Pd nanowires *in situ* by means of grazing-incidence transmission small- and wide-angle x-ray scattering (GTSAXS and GTWAXS), x-ray fluorescence (XRF) and 2-dimensional surface optical reflectance (2D-SOR). This shows how electrodeposition and the hydrogen evolution reaction (HER) compete and interact during Pd electrodepositon. During the bottom-up growth of the nanowires, we show that β-phase Pd hydride is formed. Suspending the electrodeposition then leads to a phase transition from β- to α-phase Pd hydride. Additionally, we find that grain coalescence later hinders the incorporation of hydrogen in the Pd unit cell. GTSAXS and 2D-SOR provide complementary information on the volume fraction of the pores occupied by Pd, while XRF was used to monitor the amount of Pd electrodeposited.

## 1 Introduction

Hydrogen represents a promising alternative to fossil fuel and a clean energy carrier in stationary and mobile applications[1–3]. The benefits of a successful hydrogen-based economy include the integration of intermittent renewable energies, sustainability and energy security. Remarkable effort and progress towards safe storage, sensing and production of hydrogen has been made over the last years, with the aim of realizing a hydrogen-based economy[4]. Particularly, metal hydrides and combinations of different hydrogenating metals (such as Pd and V[5], Pd and ZrCo alloys[6], Pd and Ag[7]) have proved to be a promising class of materials for hydrogen storage, since storing hydrogen in the solid form addresses the safety concerns linked to liquid or gaseous hydrogen[8,9]. Palladium is especially well-known for its catalytic activity towards hydrogen dissociation and for its ability to absorb large quantities of hydrogen, leading to the formation of α- and β-phase hydrides[10–13]. Furthermore, the hydrogenation of Pd plays a crucial role in the selectivity response of alkynes into alkenes[14]. During the hydride formation, hydrogen atoms occupy the empty octahedral sites in the face-centered cubic (fcc) structure of Pd, causing a volume expansion of the unit cell. Linear correlations exist between the lattice parameter and the amount of absorbed hydrogen[10]. Under electrochemical conditions, Pd electrodes promote the hydrogen evolution reaction (HER)[15]. It has been shown that the hydrogen produced during the HER can be absorbed by the electrode itself and lead to formation of Pd hydride[16,17].

The application of Pd-based nanomaterials to hydrogen storage and sensing is a promising emerging field[18–20]. Pd nanomaterials show better performances in terms of surface-to-volume ratio and hydride stability[21]. For instance, ordered arrays of Pd nanowires, fabricated by the template method in nano-porous anodic aluminium oxide (NP-AAO), have been successfully used in hydrogen-sensing devices with high sensitivity[22]. In previous research, we have developed a method for the synthesis of such arrays of Pd nanowires and we characterized them by *ex situ* x-ray diffraction and electron microscopy[23]. The self-arranged, highly-packed, honeycomb structure of NP-AAO and the bottom-up electrochemical growth of Pd make such material easily scalable for manufacturing purposes[24,25].

However, the interplay between structure and properties of such nanomaterials, with an emphasis on hydride formation, has only recently drawn attention. Findings about hydrogen intercalation at grain boundaries of nanocrystalline Pd[26,27] and reports of size-dependent solubility and kinetics of hydrogen absorption and desorption[28] (referred to as hydride *breathing*[29]) in Pd nanoparticles suggest that the size of the crystallites plays an important role in the dynamics of hydride formation and stability.

In this work, we investigated *in situ* the structural evolution in electrochemically grown Pd nanowires in NP-AAO templates, by means of synchrotron high-energy x-ray scattering and fluorescence. Due to their low attenuation in matter, high-energy x-rays are a powerful tool for the study of encapsulated nanostructures in NP-AAO templates. The superior photon flux of synchrotron radiation was necessary to capture real-time information. In addition, the bottom-up growth was simultaneously followed by means of 2D-SOR, which has been used in previous research to bridge atomic and macroscopic scale observations[30–33].

## 2 Material and methods

### 2.1 Template preparation

Two top-hat-shaped samples of polished polycrystalline Al (99.999%), purchased from Surface Preparation Laboratory (SPL, Netherlands), were used for the fabrication of the NP-AAO templates by means of two-steps anodization[24,34,35]. The samples were mounted in an electrochemical holder to expose only their surfaces to the anodizing solution. The two-steps method consists of a first long anodization where the oxide progressively arranges in ordered pores, followed by selective removal of the oxide which results in an Al substrate patterned by hemispherical nano-concaves. A second anodization step is subsequently performed, where the order of the pores is guided by the pre-existing nano-concaves.

The glassware used in this work were cleaned with an acid mixture of 96% $H_2SO_4$ and 65% $HNO_3$ in proportion 1:1 and then rinsed extensively with ultra-pure water (18.2 MΩ·cm). All the electrolyte solutions were prepared using ultra-pure water.

To obtain samples with different pore diameters, two kinds of anodizing solutions were used: 0.3 M oxalic acid and 0.3 M sulphuric acid. For both samples, the electrolyte temperature was lowered to 0 °C using a jacketed electrochemical cell (Gamry MultiPort Corrosion Cell) connected to a refrigerated recirculating oil bath (Julabo). The anodization in oxalic acid was conducted at 40 V while the anodization in sulphuric acid was conducted at 25 V, both using a bi-polar power supply (Kikusui PBZ-20-20). The first anodization lasted 10 h and the oxide was selectively removed in an etching bath of 0.185 M $H_2CrO_4$ and 0.5 M $H_3PO_4$ for 12 h at room temperature for each sample.

The second anodization was conducted in the same conditions of electrolyte, temperature and potential, for 30 min. Immediately after the second anodization, a sequence of electrochemical barrier layer thinning was initiated. It consisted in ramping down the applied anodizing potential to decrease the thickness of the insulating barrier layer of oxide at the pore bottom, in order to enable the electrodeposition of metal in the pore. Barrier layer thinning for the sample anodized in oxalic acid consisted of ramping down the potential from 40 V to 20 V in a time span of 30 min and again from 20 V to 1 V over 40 min, while for the sample anodized in sulphuric acid, the potential was decreased from 25 V to 1 V over 30 min. This procedure results in a template thickness of approximately 2 μm and in the formation of Y-shaped branches at the pore bottom, which affects the morphology of the deposited metal in the next step of fabrication[23,25]. The resulting pore diameter of the two templates is 40 nm for the sample anodized in oxalic acid and 25 nm for the sample anodized in sulphuric acid. The two samples are from here on referred to as template D40 and template D25, respectively.

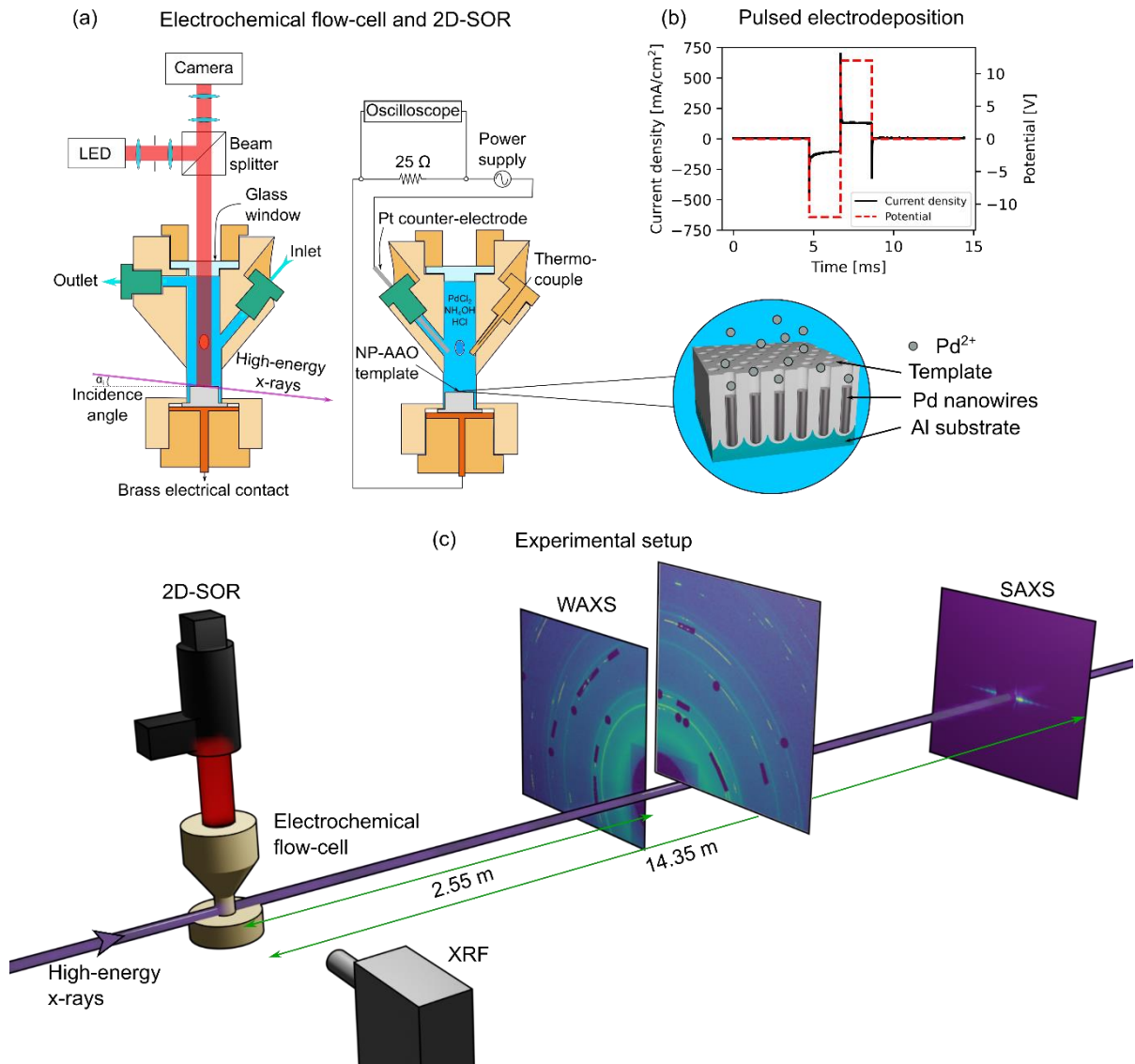

**Figure 1.** (a) Schematic drawing of the polyether ether ketone (PEEK) electrochemical flow-cell[31]. The optical aperture with the glass window enables 2D-SOR measurements. The high-energy x-rays pass through the thin PEEK walls (100 μm) of the cell and illuminate the sample at a grazing-incidence angle $α_i$. The Pd electrolyte is flown through the cell with a peristaltic pump. While pulses of potential are applied between the working and the counter electrode, the voltage across a shunt resistor of 25 Ω is measured, to estimate the current. (b) Current density measured during a typical electrodeposition potential square-wave-shaped pulse. (c) Schematic drawing of the setup used for the experiment in the P21.2 beamline at Petra III (DESY, Hamburg). GTSAXS, GTWAXS, 2D-SOR and XRF were measured simultaneously.

## 2.2 Electrodeposition

A volume of 1 L of aqueous solution (with ultra-pure water) was prepared mixing 5 mM $PdCl_2$ with 0.74 M HCl and, upon full dissolution of $PdCl_2$, 0.81 M $NH_4OH$ was slowly added. The pH was then adjusted to a value of 7.34 by small additions of $NH_4OH$. Achieving pH neutrality is necessary as alumina is stable only in the pH range 4-8[36].

The PEEK electrochemical flow-cell drawn in Fig. 1 (a), which was designed and previously used for *in situ* x-ray measurements[31,32,37–39], was employed for the experiment. In this cell, the anodized top-hat aluminium acted as the working electrode in a two-electrode setup and a Pt counter-electrode was employed. The Pd-containing electrolyte was flown into the cell using a recirculating peristaltic pump. A thermocouple in a

PEEK sleeve, inserted in the flow-cell, was used to monitor the temperature during the electrodeposition. Before the experiment, the cell and the tubing were cleaned by flowing 20% $HNO_3$ for 2 h. While cleaning, the sample and the counter-electrode were replaced by a PEEK top-hat shaped dummy sample and a Teflon stopper, respectively. After this cleaning procedure, the cell and tubing were flushed with 2 L of ultra-pure water.

The pulse electrodeposition method was employed, which consists in applying single AC square-wave pulses of electric potential (2 ms for the cathodic half-wave and 2 ms for the anodic half-wave), followed by a period of time (196 ms) where the applied potential is zero. The amplitude of the square-wave was 24 V (peak-to-peak). The benefits of this method, with regards to electrodeposition in NP-AAO, are well described in literature[25,40,41]. The Kikusui PBZ-20-20 power supply was used to program and perform the electrodeposition. A shunt resistor of 25 Ω was added in series with the electrochemical cell and the potential across it was measured with a digital oscilloscope (Picoscope PS3203D). This was done in order to estimate the current using Ohm's law. Every 7 s, the oscilloscope measured the voltage across the shunt resistor for 200 ms with a time resolution of 144 ns. A representative pulse measured during the electrodeposition is plotted in Fig. 1 (b).

Sequences of pulsed electrodeposition with a duration of 2 min, followed by a rest time of 3 min, were conducted for 60 min for the template anodized in oxalic acid and 10 min for the sample anodized in sulphuric acid. The rest sequences were added to observe the release of hydrogen from the Pd crystal lattice and at the same time to reduce the possibility of damage to the NP-AAO template by dielectric breakdown events[42]. For a systematic study, the conditions of electrolyte composition and applied potential were kept the same for the two templates investigated.

### 2.3 2-dimensional surface optical reflectance

A schematic of the 2D-SOR setup, which was mounted on the diffractometer together with the electrochemical cell, is shown in Fig. 1 (c). The electrochemical flow-cell has an optical access on the top, through a glass window. The light source is a red LED with a power of 700 mW and a wavelength of 650 nm (Thorlabs M625L3). The LED light is first reflected by the sample and then by a beam splitter to a 2D CMOS camera with (Allied Vision). A detailed description of the 2D-SOR setup, used in combination with electrochemical measurements, was previously published[31,32].

### 2.4 *In situ* synchrotron measurements

The synchrotron measurements were performed at the Swedish Material Science beamline P21.2 (Petra III, DESY, Hamburg). The structural evolution of the electrodeposited Pd was followed by means of small- and wide-angle x-ray scattering ( GTSAXS and GTWAXS), and x-ray fluorescence (XRF). The grazing-incidence transmission geometry was employed to study the embedded nanostructures while reducing distortion phenomena that usually affect grazing-incidence SAXS. In fact, in the grazing-incidence transmission geometry, the contributions to the scattering amplitude that are normally described by the distorted-wave Born approximation can be neglected[43]. A 68.5 keV x-ray beam, focussed to 13 μm × 100 μm (vertical × horizontal), illuminated the sample at a grazing-incidence angle of 0.15°.

XRF was measured by an Amptek XR-100 CdTe detector, positioned at approximately 90° relative to the direct beam. The energy channels of the XRF detector were calibrated using standards of Ag, Cu, Mo and Rb. XRF spectra were collected during the electrodeposition and the Pd Kα emission line was subsequently integrated. The Pd signal from the electrolyte only, right before the electrodeposition, was used as baseline.

GTSAXS was measured by a Varex XRD 4343CT flat panel detector (pixel size 150 μm) positioned at a distance of 14.35 m from the diffractometer centre, as determined by calibration using a silver behenate reference. An evacuated flight tube was positioned in between the diffractometer centre and the GTSAXS detector. GTWAXS was measured by two additional Varex XRD 4343CT detectors positioned at a distance of 2.55 m. Fig. 1 (c) is a depiction of the experimental setup showing how the detectors were positioned. Where necessary, the strong Bragg reflections arising from the polycrystalline Al substrate were attenuated by some tungsten beam stoppers and lead foils, attached to a plexi-glass panel in front of the detectors. The detector distances and their tilts were calibrated using NIST LaB$_6$ as standard reference material (measured in transmission geometry) and *pyFAI* as the calibration software[44–46]. A description of the calibration procedure and the sample alignment, with an emphasis on the prevention of artefacts due to sample misplacements, can be found in Appendix A. We followed the first 10 min of deposition, while GTSAXS, GTWAXS, XRF and 2D-SOR were collected continuously with an exposure time of 5 s. For sample D40, the electrodeposition and the measurement continued for 1 h, recording one image every 10 s (with an exposure time of 5 s). To rule out any beam-induced effect, we measured the sample at two different positions, rotating the sample around its surface normal back and forth by 5° between an image acquisition and the following one.

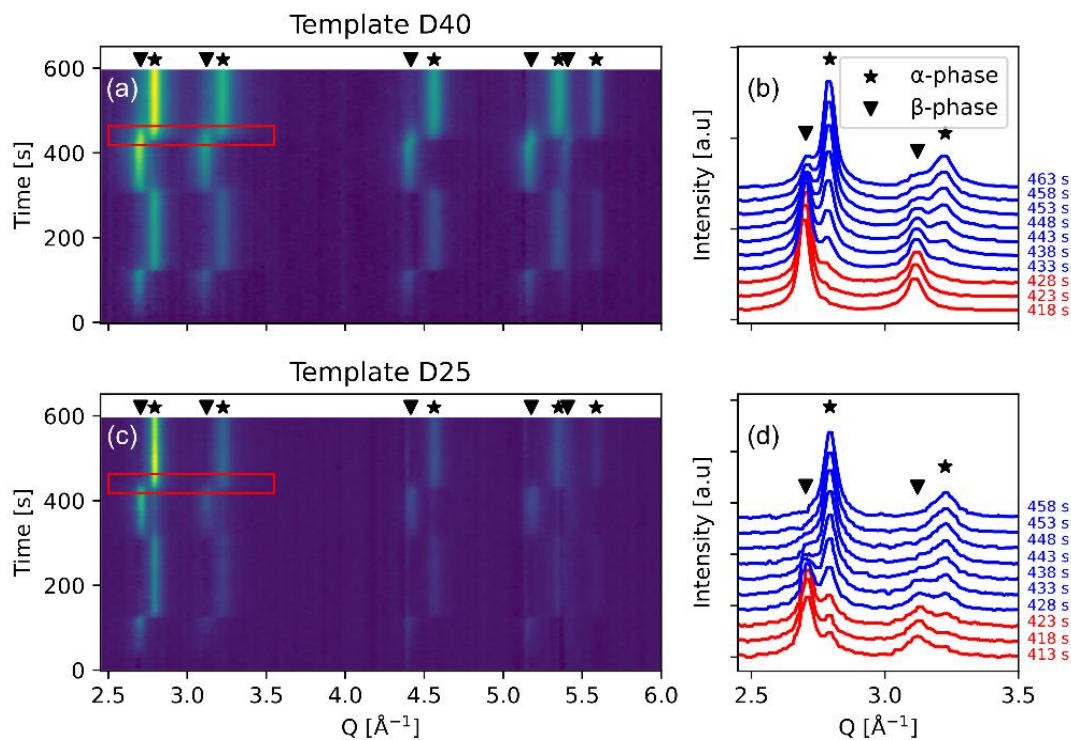

**Figure 2.** Waterfall plots of the GTWAXS patterns as function of time, for the Pd electrodeposition in template D40 [(a) and (b)] and in template D25 [(c) and (d)]. The boxed areas in (a) and (c) are reported in (b) and (d), respectively. The red lines were acquired while the electrodeposition was enabled and the blue lines were acquired when the electrodeposition was interrupted (electrochemical cell left at open circuit potential).

## 3 Results and discussion

Fig. 2 shows an overview of the GTWAXS data collected during the electrodeposition of Pd in template D40 [(a) and (b)] and template D25 [(c) and (d)]. Every horizontal line of the waterfall plots in Fig. 2 (a) and (c) is a diffraction pattern obtained from the integration of the GTWAXS detectors over the whole accessible azimuthal range. The reader is referred to Appendix B for a more detailed description of the data processing. Qualitatively, two fcc structures have been found, which were identified as α- and β-phase PdH$_x$. However, the lattice constant of these two phases is proportional to the amount of hydrogen absorbed in the fcc unit cell. As a reference, we marked the positions of the α- and β-phase PdH$_x$ calculated by using a lattice parameter of 3.895 Å for the α-phase (corresponding to a molar fraction x=0.02) and 4.025 Å for the β-phase (corresponding to a molar fraction x=0.58). The plots in Fig. 2 (b) and (d) were extracted from the boxed regions in Fig. 2 (a) and (c), drawn across some phase transitions. Here, the two phases coexist for approximately 30 s, before the β-phase almost completely disappears and the α-phase becomes dominant. What causes the phase transition is the interruption of the electrodeposition. As described in Section 2, we alternated sequences of deposition lasting 2 min with sequences of rest lasting 3 min. The red lines in Fig. 2 (b) and (d) were acquired during the electrodeposition, while the blue lines were acquired after the electrodeposition was suspended.

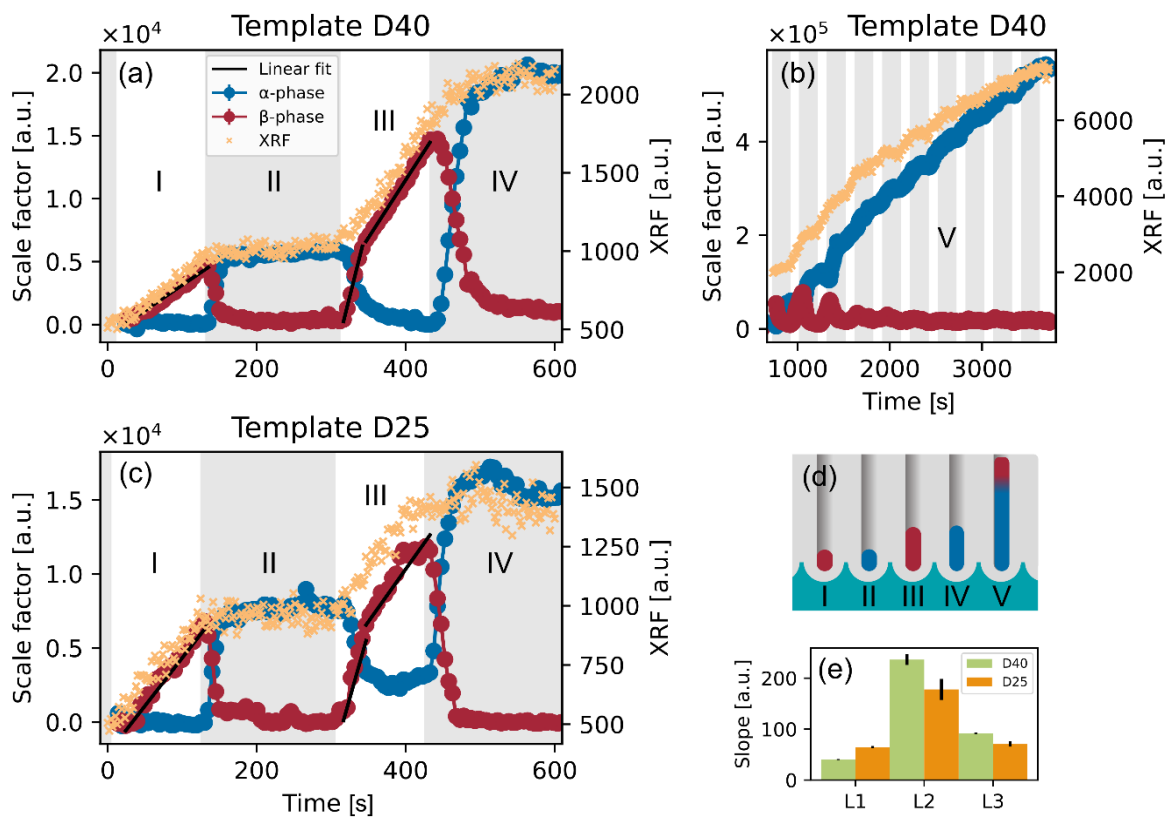

**Figure 3.** X-ray fluorescence and scale factors extracted from the Rietveld refinement of GTWAXS patterns collected *in situ* during the electrodeposition of Pd in template D40 [(a) and (b)] and in template D25 (c). (d) Cartoon showing the Pd nanowires at stages I to V: the blue segments correspond to α-phase hydride and the red segments correspond to β-phase hydride. (e) Slope of the black lines in (a) and (c) in order of appearance from left to right.

The one-dimensional GTWAXS patterns were fed into a sequential Rietveld refinement algorithm, to obtain quantitative structural information. Appendix B contains more information on the refinement procedure, as well as a representative GTWAXS frame and its refinement (see Fig. B.2). In Fig. 3 (a), (b) and (c) the scale factors of the two phases, extracted from the Rietveld refinement, are reported. The scale factor is a measure of the contribution of a certain phase to the overall intensity of a diffraction pattern. The grey areas in Fig. 3 (a), (b) and (c) correspond to periods of time where the electrodeposition was interrupted. We identified four stages, namely I to IV, were the two templates show a similar evolution of the scale factors. For sample D40, we identified a fifth stage (V). In region I, the scale factor of the β-phase increases linearly, while the scale factor of the α-phase is zero. This means that the nanoparticles nucleating and growing in the first 120 s of deposition form as Pd hydride rather than pure Pd. This phenomenon can be attributed to HER, happening in parallel with the electrodeposition, while the applied potential is cathodic. In region II, the scale factor of the β-phase decays (in a time span of 30 s), while the α-phase appears. Here, the power supply is suspended and both the electrodeposition and the HER are stopped. The hydrogen absorbed in the crystal lattice diffuses out of the nanowires, causing a phase transition. In region III, the scale factor of the α-phase decays while the scale factor of the β-phase increases. The increase of the β-phase scale factor can be fit by two adjacent lines. In the first one, the increase is rapid and can be attributed to the conversion of the pre-existent α-phase hydride into β-phase hydride. The second, slower increase can be attributed to an increasing quantity of Pd in the pores due to the electrodeposition. Region IV can be described by a phase transition from β- to α-phase hydride, similarly to region II. For template D40, where the electrodeposition continued for 3600 s, we identified a fifth stage (V). Here, the formation of α-phase hydride becomes dominant and the formation of β-phase decreases over time. We surmise that only a fraction of the nanowires at the metal/electrolyte interface becomes β-phase hydride and that part of the absorbed hydrogen diffuses either towards the electrolyte and towards the bottom of the nanowire, forming the α-phase hydride. The hydride phase observations are supported by findings of cathodic current decrease during the deposition sequences, as described in greater detail in Appendix C.

It is possible that the hydrogen diffuses quite rapidly from the metal/electrolyte interface to the bulk of the nanowire. The diffusion coefficient is well-described by the Arrhenius equation

$$D = D_0 \exp(-\frac{E_D}{RT}),$$

where $D_0$ is a pre-exponential factor, $E_D$ is the activation energy, R is the gas constant and T is the absolute temperature. The reported values found in literature for $D_0$ and $E_D$ are $2.4 \times 10^{-7}$ m$^2$/s and 21.1 kJ/mol, respectively[47]. At room temperature, the Arrhenius equation yields a diffusion coefficient of hydrogen in Pd of $4.804 \times 10^{-11}$ m$^2$/s. Using the equation for the root mean square displacement due to Brownian motion, $x=\sqrt{2Dt}$, it can be estimated that hydrogen takes approximately 10 ms to diffuse across a 1 μm long nanowire.

In parallel with the scale factors, the XRF signal arising from the Pd Kα emission line is reported. The XRF is insensitive to hydride formation and is only proportional to the amount of Pd atoms inside the pores. The XRF signal only increases in stages I, III and the white areas in stage V, while it is constant in regions II, IV and the grey regions of stage V. This is in agreement with the fact that the nanowires only grow while an electric potential is supplied and supports the evidence that the phase transition in the grey areas is due to material pre-existing as β-phase hydride. A visual description of stages I to V, representing the growth and the phase transitions is shown in Fig. 3 (d).

For comparison, we reported the slopes resulting from the linear fits of the β-phase scale factors in Fig. 3 (e). Initially, the growth is faster in template D25. However, in stage III, the conversion of the α-phase into β-phase and the electrodeposition rate are more rapid in template D40. This could be due to the higher porosity of template D40 (14%, against 12% porosity of template D25) and therefore to the larger availability of surface area that is actively involved in HER.

The lattice parameters extracted from Rietveld refinement, are plotted in Fig. 4. The discontinuities in Fig. 4 (a) and (c) are due to the fact that the phases are not simultaneously detected in the first 600 s of electrodeposition. The lattice parameter is a measure of the amount of hydrogen absorbed in the unit cell[48]. As a reference, the lattice parameters of pure Pd, α-phase PdH$_{0.02}$ and β-phase PdH$_{0.58}$ are reported. For both phases, the lattice parameter is greater in template D40 than in template D25. At the beginning of each deposition sequence (regions I and III), the lattice parameter increases, as the Pd absorbs an increasing quantity of hydrogen and the α-phase undergoes a phase transition to β-phase. Similarly, when the deposition is interrupted, the lattice parameter decreases, as the lattice releases hydrogen by diffusion. Fig. 4 (e) schematically represents how the increasing occupancy of the octahedral sites of the Pd fcc structure by H atoms leads to the volume expansion of the unit cell and the formation of the β-phase hydride. In previous research, we observed an anisotropic strain affecting the lattice constant[23,39]. However, in this experiment we could not determine any deviatoric strain (i.e., that after subtracting hydrostatic strain) within the experimental uncertainty.

Fig. 4 (b) shows the lattice constant in template D40, as the growth proceeds up to 1 h. Here we can distinguish two subdivisions in stage V. In stage $V_A$, the lattice parameter fluctuates around the predicted value for the α-phase PdH$_{0.02}$: it increases during the deposition sequences and it decreases during the rest periods. In stage $V_B$, however, the lattice parameter decreases below the value predicted for the molar fraction x=0.02 and tends towards the lattice parameter of metallic Pd (x=0). Since in stage $V_B$ the variation in lattice parameter is subtle and its corresponding molar fraction variation is small (from 0.02 to 0), the distinction between α-phase hydride and metallic Pd is no longer measurable. Based on this analysis of the lattice parameter, it should be noticed that the scale factor of the α-phase hydride reported in Fig. 3 (b) is not entirely representative of a hydride phase but rather of a combination of hydride and pure Pd, especially after t=1480 s.

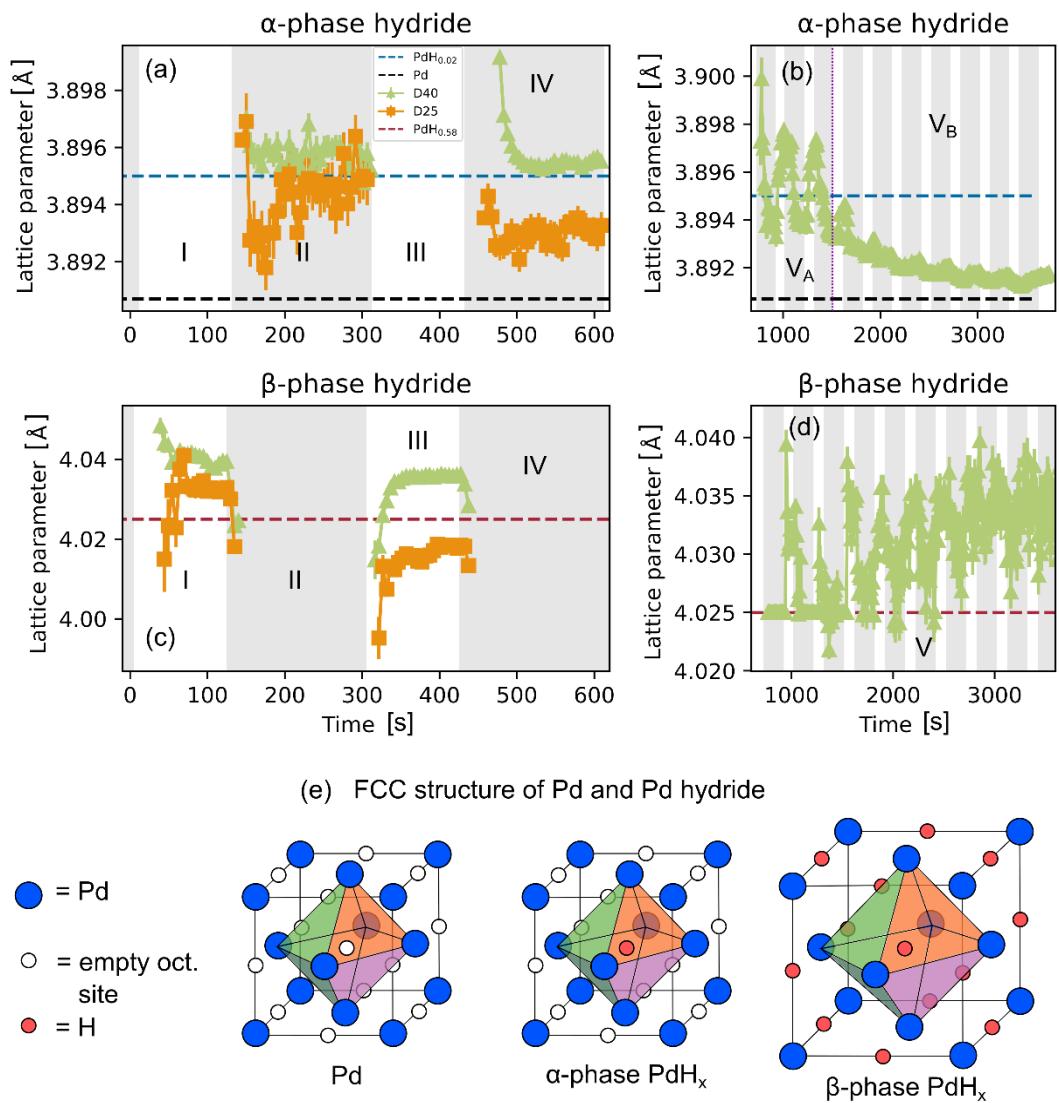

**Figure 4.** Lattice parameters extracted from Rietveld refinement of GTWAXS patterns of the α- [(a) and (b)] and β-phase hydride [(c) and (d)]. The theoretical lattice parameter of $PdH_{0.02}$ (α-phase) $PdH_{0.58}$ (β-phase) and pure Pd are marked with blue, red and black lines, respectively. (e) Atomistic model of the Pd fcc structure upon hydrogenation.

On the other hand, the lattice parameter of the β-phase in Fig. 4 (d) remains well above the value for $PdH_{0.58}$. The scale factor in Fig. 3 (d), however, showed that the β-phase formation is less prominent towards the end of the electrodeposition. This strengthens the argument that there is a small region of the nanowires, in proximity of the metal/electrolyte interface, which consists of β-phase hydride, while the rest of the nanowire resembles α-phase hydride and, progressively, pure Pd. This model is compatible with the fact that after the electrodeposition, and after the sample is removed from the electrochemical environment, our *ex situ* measurements performed in previous research revealed no presence of Pd hydride[23].

From the Rietveld refinement, the size of the crystallite was extracted and reported in Fig. 5 (a) and (b) for template D40 and (c) for template D25. In Fig. 5 (a), by the end of stage I, the crystallite size decreases.

This is due to the decrease in scattering domain size of the β-phase hydride as the hydrogen leaves the crystal lattice, leading to the transition into α-phase hydride. Conversely, at the beginning of stage III in Fig. 5 (c), there is a sudden increase in crystallite size, from 8 to 12 nm, although the pre-existing α-phase had a crystallite size of 12 nm. This can be explained similarly as an increase of scattering domain size of the β-phase during the HER-initiated transition.

Overall, in both templates, there is an increase in crystallite size from approximately 9 to 18 nm (and above 20 nm for template D40). This is attributed to the coalescence of nanocrystals, schematically represented in Fig. 5 (d). Coalescence phenomena are not unusual under electrochemical conditions. For instance, the electrochemical Ostwald ripening in nanocrystalline metal ensembles has been reported[49].

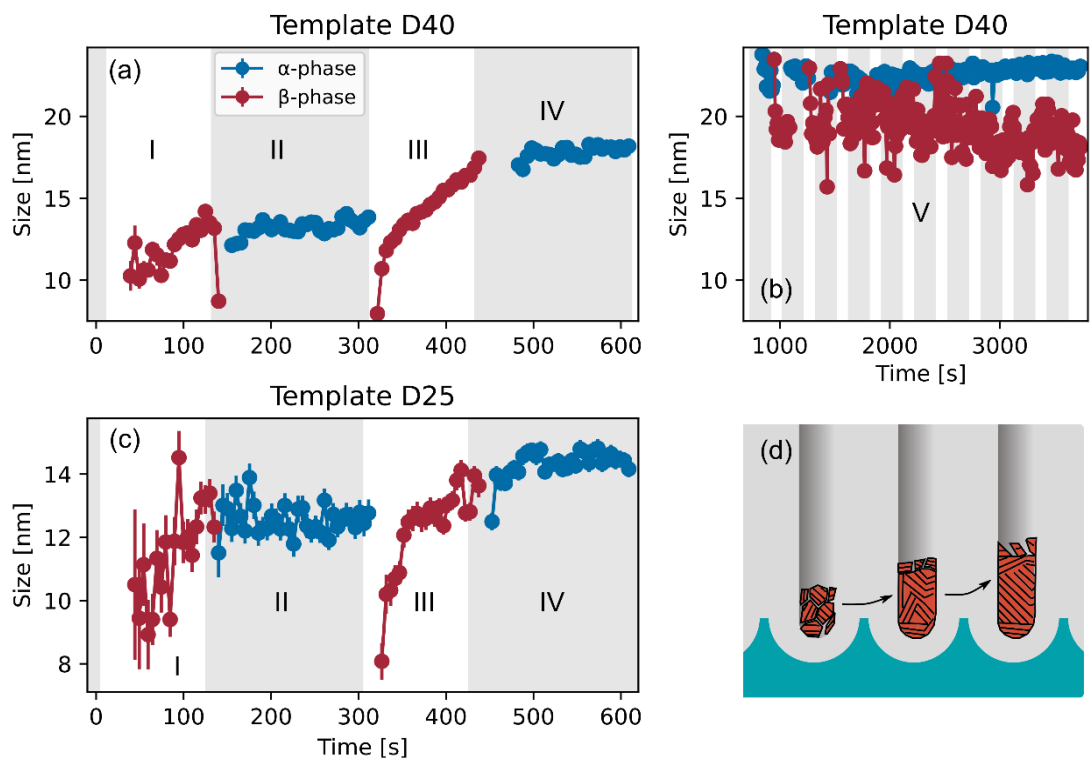

**Figure 5.** Crystallite size extracted from the Rietveld refinement of GTWAXS patterns measured during the electrodeposition of Pd in template D40 [(a) and (b)], and in template D25 (c). Schematic depiction of grain growth and aggregation during the electrodeposition (d).

This grain coalescence could be the cause of the reduced absorption of hydrogen after long deposition times (stage V). It is known that the intercalation and the absorption of hydrogen is mediated by grain boundaries[26]. In addition, hydrogen absorption proceeds in three stages: $H_2$ molecules are dissociated at the Pd surface, subsequently hydrogen penetrates into subsurface sites and then diffuses into the bulk[50]. The lower surface-to-volume ratio of the nanocrystals, caused by grain coalescence, potentially leads to in a reduced availability of absorption sites and thus to a lower rate of hydride formation.

Simultaneously, GTSAXS and 2D-SOR were measured, which revealed complementary information, available in Appendix D, on how the deposition proceeds first in the dendritic branches of the nano-pores and later in the body of the pore.

### 4 Conclusion

*In situ* GTWAXS measurements revealed that electrochemically fabricated Pd nanowires in NP-AAO templates grow as $PdH_x$ in the early stages of the electrodeposition. Alternating sequences of electrodeposition with sequences of rest causes transitions between β- and α-phase hydride, respectively.

We proposed a model where the hydrogen, produced by HER at the working electrode, is absorbed by the Pd nanowires, forming a small region of β-phase hydride close to the metal/electrolyte interface. The hydrogen rapidly diffuses to the bulk of the nanowires, that are less populated by hydrogen and therefore have a lower partial pressure, to form an α-phase hydride, which at later times becomes pure Pd as the hydrogen degresses. At later deposition times, the coalescence of Pd nanocrystals in the pores causes lower rates of hydrogen absorption, which is mediated by grain boundaries.

The correlation between the hydride formation and the deposition/rest sequences adopted in this work might be an interesting tool to control the hydrogen intake in the Pd nanowires. This might be useful in applications where the hydrogen content enables the selectivity of Pd towards some specific reactions, e.g., the conversion of alkynes into alkenes[14]. In addition, we demonstrated that hydrogenating metals in nano-porous alumina are a promising class of composite materials for the simultaneous production and storage of hydrogen in the solid form. In the future, a combination of Pd with less expensive hydrogenating metals, such as V or Ni can be investigated towards this particular application.

# Appendix A

**Calibration and sample alignment**

For the evaluation of absolute lattice parameters, the experimental setup was calibrated by a reference powder sample. It is critical that the effective distances of the reference powder and real sample to the detector are identical. This was achieved by aligning both samples to the vertical rotation axis of the diffractometer and assuring that the incoming beam hit the sample at its centre under grazing incidence. The detailed alignment procedure is described in the following.

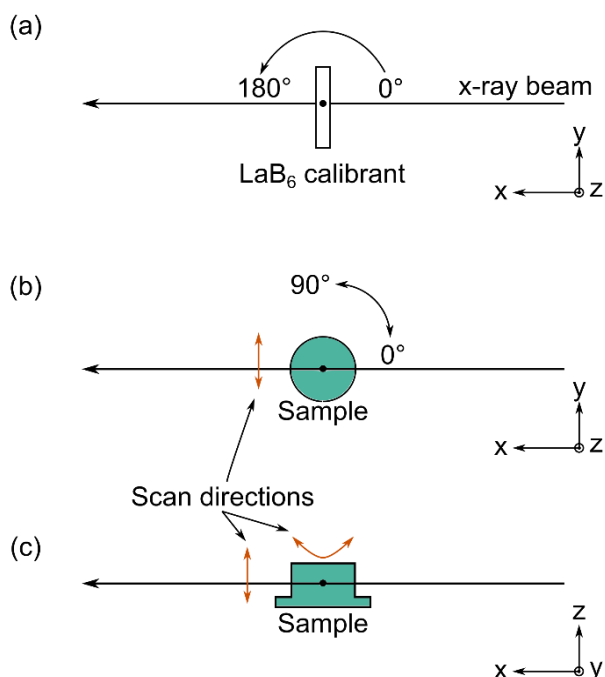

**Figure A.1.** (a) Top-view of a LaB$_6$ calibrant aligned in the centre of the diffractometer and measured in the transmission geometry; (b) top-view and (c) side-view of the top-hat shaped sample, scanned in the x-ray beam during the alignment.

First, the incident x-ray beam was aligned to the intersecting rotation axes of the diffractometer around the z- and y-axes (the latter was used to set the grazing incidence angle). A disk of LaB$_6$ standard reference powder was prepared by pressing the powder in a metal washer between two foils of Kapton tape, mounted on a spinner (to improve the homogeneity of the diffraction rings). The WAXS arising from the reference sample was measured in the transmission geometry [see Fig. A.1 (a)]. In order to ensure that the reference coincided with the z-rotation axis, we rotated it about the z-axis and measured WAXS at two positions: 0° and 180°. We then fed the measured WAXS patterns into pyFAI to calculate the sample-to-detector distance (SDD) for the two positions. We then translated the sample along the x-axis to compensate for the difference in the SDD and measured another WAXS pattern. This last measurement was used to calibrate the setup and to obtain the coordinates of the point of normal incidence, to perform the azimuthal integration of all the subsequent GTWAXS patterns measured in this work.

The top-hat shaped samples used in this work were aligned against the x-ray beam to be in the same position as the reference material, i.e., the z-rotation axis of the diffractometer. The alignment procedure consisted in (i) scanning and centring the sample across the beam in the y-direction [see Fig. A.1 (b)] for two rotation angles about the z-axis: 0° and 90°; (ii) scanning the sample along the z-axis and centring the sample at the beam cut-off height [see Fig. A.1 (c)]; (iii) scanning the tilt angle and rotating the sample to the position where the maximum beam transmission is measured. This last step was reiterated for the z-axis rotations 0° and 90°. Before setting the grazing incidence angle, the sample was translated again along the z-axis to the cut-off height. Furthermore, we monitored the sample position with an x-ray eye to ensure that the sample height was not subject to variations due to misplacements of the horizontal rotation axis against the x-ray beam.

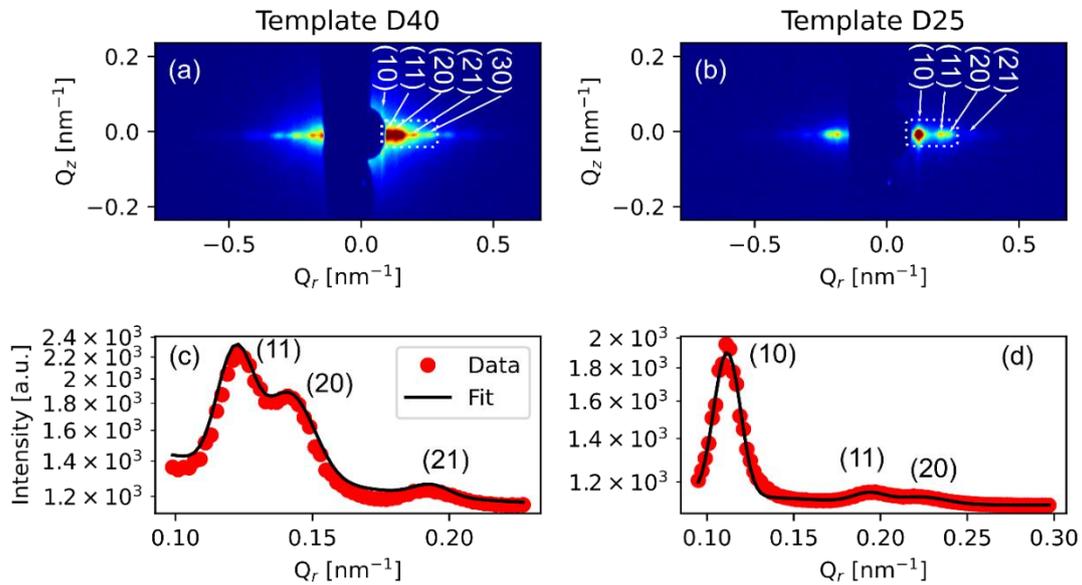

**Figure B.1.** Representative GTSAXS patterns collected during the electrodeposition of Pd in template D40 (a) and D25 (b); 1-dimensional patterns extracted from the boxed regions [(c) and (d)], plotted together with their fits.

# Appendix B

**GTSAXS and GTWAXS data processing**

The 2-dimensional GTSAXS patterns were integrated in the boxed area shown in Fig. B.1, to obtain 1-dimensional patterns. The peaks in the GTSAXS pattern were assigned a set of two (h,k) Miller indices, as the NP-AAO can be described by a two-dimensional hexagonal lattice[51]. The 1-dimensional GTSAXS patterns were fitted by a multiple-peak fitting algorithm, based on the Python module *lmfit*[52], where the peak positions were given as a starting guess based on the Miller indices (h,k) and the inter-pore distance $a$ through the following equation:

$$Q_{hk} = \frac{4\pi\sqrt{h^2 + hk + k^2}}{\sqrt{3}a}$$

which describes the peak positions for a two-dimensional hexagonal lattice. The peaks were fitted with Voigt profiles to return the peak widths of the GTSAXS pattern and the background was fitted by Chebyshev polynomials with 3 coefficients. The peak widths were used to estimate the porous domain size using Scherrer's formula. Fig. B.1 (c) and (d) are an example of integrated GTSAXS patterns and their fit.

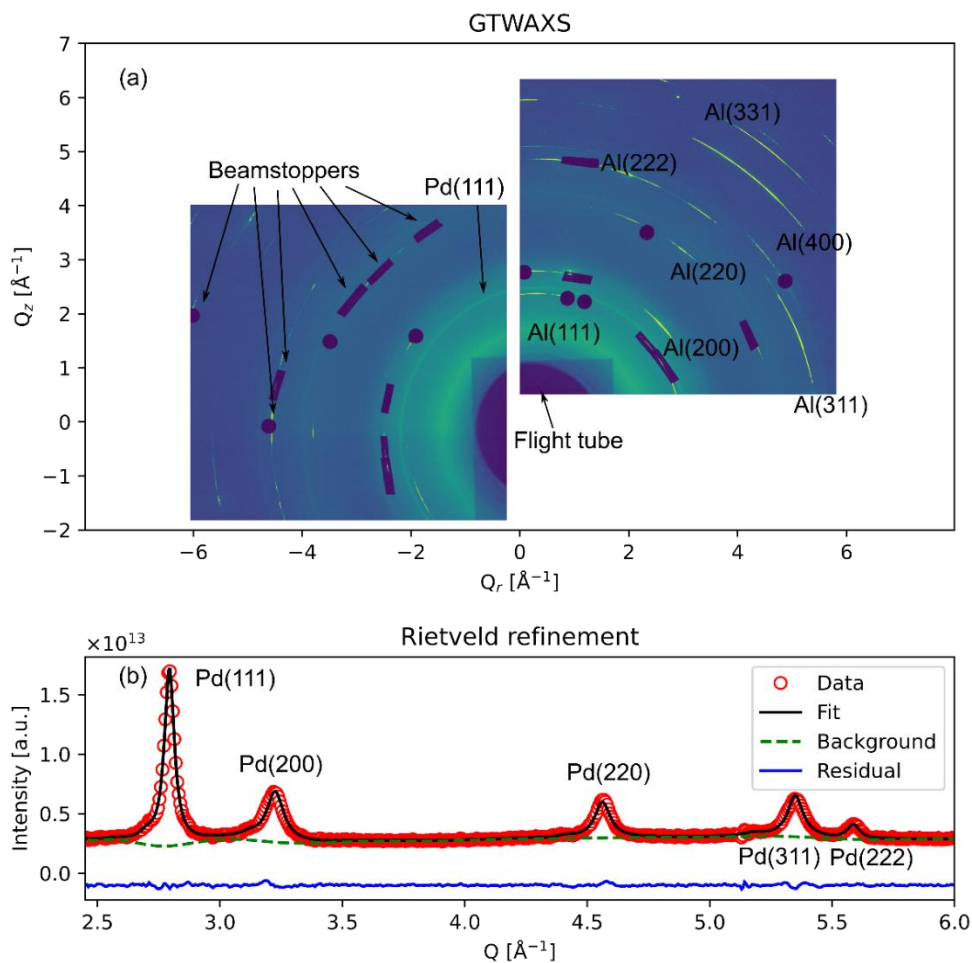

**Figure B.2.** A representative GTWAXS pattern recorded during the electrodeposition of Pd in the template anodized in oxalic acid (a). Azimuthal integral of the GTWAXS images, together with its Rietveld refinement (b). The Al reflections and the beam stoppers were masked out from the integration.

The 2-dimensional GTWAXS patterns were integrated using the *pyFAI* "Multi-geometry" module, which can take as an input images collected from different detectors at the same time[45]. The areas containing the beam stoppers and the Al reflections were masked out from the integration. The GTWAXS images measured prior to the electrodeposition were used as background. The GTWAXS patterns were integrated over the whole accessible azimuthal range, and in a radial range from 2.5 to 6 Å$^{-1}$, in Q-units, with a total number of radial bins of 600.

The GTWAXS 1-dimensional pattern of the LaB$_6$ reference was loaded in GSAS-II[53] and the peaks were fitted to obtain the instrumental resolution function (a calibration file containing the Caglioti coefficients[54–56], amongst other instrumental parameters). These parameters were loaded in GSAS-II and kept fixed for every *in situ* pattern.

The *in situ* 1-dimensional GTWAXS patterns were fed to a sequential Rietveld refinement algorithm in GSAS-II. The intensity scale factor, the lattice parameter and the crystallite size of each phase were refined. The background of the patterns was fitted by a Chebyshev polynomial with 12 coefficients. The sequential refinement was performed in the reverse order, starting from the last acquired GTWAXS pattern, and the output parameters of every refinement was used as the initial guess of the following refinement. Fig. B.2 shows a typical 2-dimensional GTWAXS pattern, an integrated 1-dimensional pattern and the corresponding Rietveld refinement. The intensity along the Pd diffraction rings is homogeneous, which is a consequence of the isotropic orientation of the crystallites.

# Appendix C

**Cathodic current**

The mean cathodic current flowing in the electrochemical cell during the electrodeposition is reported in Fig. C.1 [(a) and (b)] for template D40 and (c) for template D25. During every deposition sequence, the current decreases, regardless of the pore size. Furthermore, the initial current of every deposition stage is always higher than the last current value measured at the end of the previous deposition stage. This current behaviour can be attributed to the fact that $PdH_x$ is a semiconductor and its resistivity is proportional to the hydrogen content[57]. During the potential-limited pulse electrodeposition, Pd incorporates an increasing quantity of hydrogen, due to HER, and its resistivity increases, leading to a current decrease. An exception to this decreasing current behaviour is in stage I of Fig. C.1 (a). Here, the current is subjected to fluctuations due to electrostatic instabilities during the initial growth of the nanowires. These instability events are known to happen in templates treated by means of barrier layer thinning and they are due to the higher conductivity of the metal deposited compared to the one of the electrolyte. In addition, the instability events are more frequent in the initial stage of growth, where the electrodeposition proceeds faster in nano-pores with a slightly thinner barrier layer than the one of other surrounding nano-pores[41].

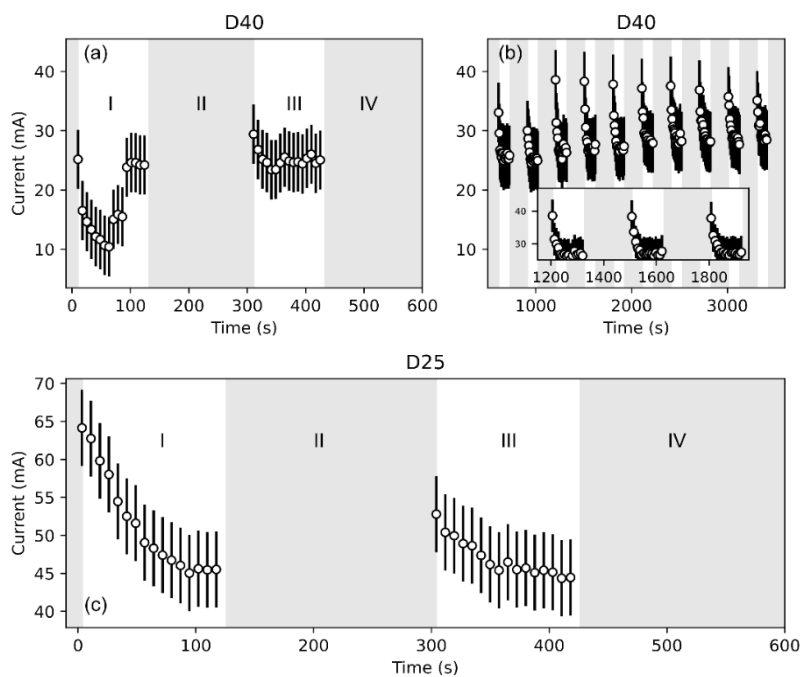

**Figure C.1.** Mean cathodic current extracted from the negative half-waves measured during the Pd electrodeposition in templates D40 [(a) and (b)] and D25 (c).

# Appendix D

**2D-SOR and GTSAXS domain size**

Some frames extracted from the *in situ* 2D-SOR measurements are reported in Fig. D.1 (a), (b) and (c) for template D40, and (d), (e) and (f) for template D25. Thanks to the spatial resolution of this technique, it was possible to visually monitor the uniformity of the reflectance over the whole surface of the template.

Quantitatively, the mean reflectance over the whole sample surface is shown in Fig. D.2 (a) and (b), where the SOR signal decreases during the deposition steps. It has been proven that metal nanoparticles in nanoporous alumina templates are subjected to plasma absorption of optical light, and that the absorbance is proportional to the volume fraction occupied by the metal particles[58]. Therefore, the reflectance decrease can be due to the increasing aspect ratio of the nanowires in the pores during the electrodeposition stages.

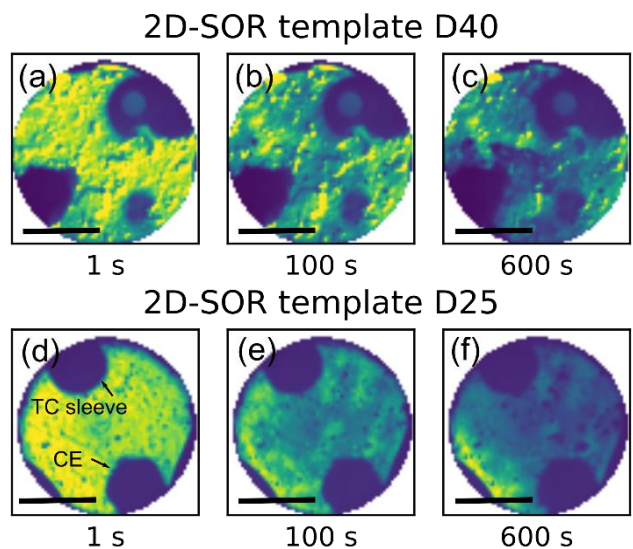

**Figure D.1.** 2D-SOR images collected during the electrodeposition of Pd in template D40 [(a), (b) and (c)] and in template D25 [(d), (e), (f)]. The PEEK sleeve containing the thermocouple and the Pt counter-electrode are also visible. The scale bar corresponds to a length of 3 mm.

It has been shown in previous research focused on the electrodeposition of Sn in NP-AAO templates that the porous domain size extracted by GTSAXS patterns decreases as the pores are filled[59]. GTSAXS and SOR can thus provide complementary information. In Fig. D.2, we reported the GTSAXS domain size (extracted from the mean peak widths of the GTSAXS profiles), together with the SOR signal. In the first 1000 s in Fig. D.2 (a) and in the 600 s of deposition in Fig. D.2 (b), the GTSAXS domain size is quite constant, while the SOR decreases more rapidly. Conversely, after approximately 1000 s, the SOR signal remains constant, while the GTSAXS domain size decreases more sharply. In research about the template-assisted growth of Ag nanowires in NP-AAO treated with barrier layer thinning (a similar template to the one used in this work)[41], two regimes of growth were identified: (i) a heterogeneous nucleation of nanocrystals occurs at the dendritic pore bottoms, preferentially in those where the barrier layer is thinner and (ii) a more uniform growth of the nanowires occurs when the branches are filled, caused by the decrease of electrochemically active surface area. This suggests that SOR is more sensitive to the initial stages of nucleation of $PdH_x$ nanocrystals in the dendritic pores bottom, while GTSAXS is more sensitive to the height increase of nanowires that have nucleated in the first stages of deposition.

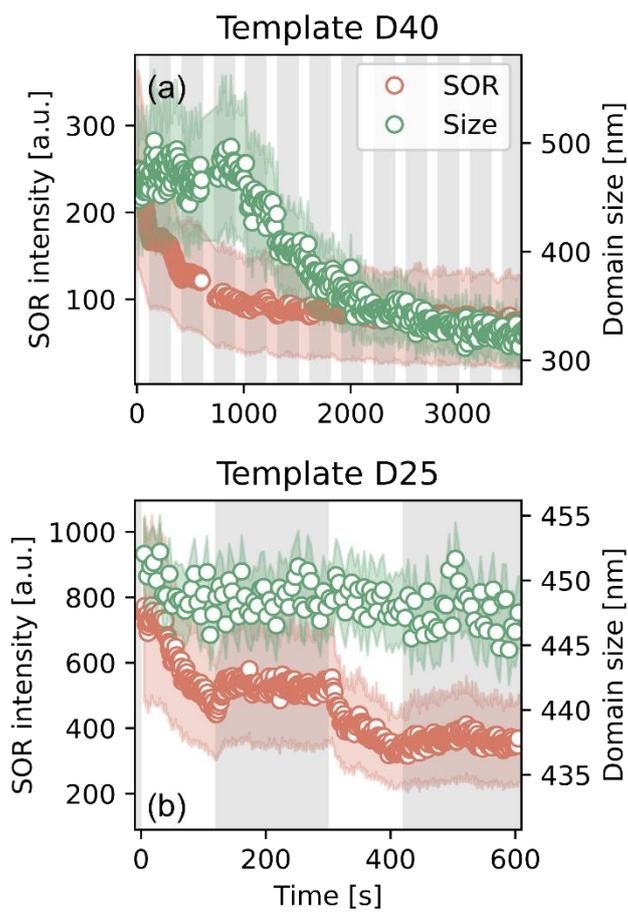

**Figure D.2.** Mean 2D-SOR and GTSAXS porous domain size as a function of time for template D40 (a) and template D25 (b).

**Acknowledgements**

We acknowledge DESY (Hamburg, Germany), a member of the Helmholtz Association HGF, for the provision of experimental facilities. Parts of this research were carried out at PETRA III at the Swedish Materials Science beamline P21.2. Beamtime was allocated for proposal I-20211146 EC. We would like to thank Sven Gutschmidt for assistance in setting up the experiment.

This work was financially supported by the Swedish Research Council through the Röntgen-Ångström-Cluster 'In-situ High Energy X-ray Diffraction from Electrochemical Interfaces (HEXCHEM)' (Project No. 2015-06092), by the 'Atomic Resolution Cluster'- a Research Infrastructure Fellow program of the Swedish Foundation for Strategic Research, and project grant 'Understanding and Functionalization of Nano Porous Anodic Oxides' (Project No. 2018-03434) by the Swedish research council. We acknowledge financial support by NanoLund.